# PlasmoID: A dataset for Indonesian malaria parasite detection and segmentation in thin blood smear

Hanung Adi Nugroho [1,*], Rizki Nurfauzi [1], E. Elsa Herdiana Murhandarwati [2], Purwono Purwono [2]

[1] Department of Electrical and Information Engineering, Faculty of Engineering, Universitas Gadjah Mada
[2] Department of Parasitology, Faculty of Medicine, Public Health and Nursing, Universitas Gadjah Mada
* Correspondence: adinugroho@ugm.ac.id; Tel.:

**Abstract:** Indonesia holds the second-highest-ranking country for the highest number of malaria cases in Southeast Asia. A different malaria parasite semantic segmentation technique based on a deep learning approach is an alternative to reduce the limitations of traditional methods. However, the main problem of the semantic segmentation technique is raised since large parasites are dominant, and the tiny parasites are suppressed. In addition, the amount and variance of data are important influences in establishing their models. In this study, we conduct two contributions. First, we collect 559 microscopic images containing 691 malaria parasites of thin blood smears. The dataset is named PlasmoID, and most data comes from rural Indonesia. PlasmoID also provides ground truth for parasite detection and segmentation purposes. Second, this study proposes a malaria parasite segmentation and detection scheme by combining Faster RCNN and a semantic segmentation technique. The proposed scheme has been evaluated on the PlasmoID dataset. It has been compared with recent studies of semantic segmentation techniques, namely UNet, ResFCN-18, DeepLabV3, DeepLabV3plus and ResUNet-18. The result shows that our proposed scheme can improve the segmentation and detection of malaria parasite performance compared to original semantic segmentation techniques.

## 1. Introduction

*Malaria* is a disease caused by protozoa parasites transmitted by a bite of an infected female Anopheles mosquito [1][2]. Malaria becomes a significant health problem in part of tropical and subtropical areas having inadequate health facilities. Africa is the most dangerous malaria area, contributing 93% of 229 million global cases and almost half a hundred thousand malaria deaths in 2019, and *P.falciparum* is the deadliest malaria [3][4].

Referring to the World Malaria Report published by World Health Organization (WHO) in 2020, Indonesia is ranked as the second country with the highest malaria rate after India in Southeast Asia. In addition, some malaria-endemic districts in Eastern Indonesia show a high concentration of positive malaria cases and a number of malaria sufferers (Annual Parasite Incidence/API) [5]. For example, based on the Ministry of Health data in 2019, Papua Province became the most significant contributor to malaria cases, up to 86% or 216,380 cases. Then, the rate is followed by East Nusa Tenggara Province with 12,909 cases and West Papua Province with 7,079 cases. The other region with high endemic areas is central Indonesia, precisely in Penajaman Paser Utara Regency, Kalimantan Timur Province [6].

Microscopic examination, a golden standard for malaria parasite diagnosis, can provide precise information about the condition of malaria patients, which others cannot offer. However, the examiner needs to screen up to 5000 red blood cells (RBC) in a thin blood smear to diagnose. Thus, the effectiveness and accuracy of the decision results are heavily independent of the examiner's experience, which was a significant limitation of the manual microscopic examination [7][8]. Malaria areas in Indonesia are remote areas with very few health facilities and very few trained and experienced microscope examiners. This limitation needs to be overcome with an automated, fast, reliable, and accurate computer-aided diagnosis (CAD) system.

There are two common approaches to segmenting parasites in thick and thin blood smear images. The first is by traditional image processing approaches, and the second is by modern deep learning approaches. The traditional image processing approaches are usually applied for noise filtering [9][10], enhancing [11][12], normalization [13], and basic segmentation [14][15][16][17][18]. The traditional image processing approaches require a deep knowledge of the target object characteristics whole the data. In addition, this approach has poor generalization in the classification task due to the model's inability to handle the inherent variability of images from different domains.

Recently, the deep learning approach has successfully solved the limitations of the traditional image processing approach and has become popular. Some recently advanced deep learning techniques for malaria localization offer good performances [19][20]. However, the result only provides the predicted box coordinates but not for the parasite shape. Malaria parasite image segmentation is significant since the output system is used for further malaria diagnosis, such as treatment recommendations.

Several studies have proposed semantic segmentation techniques for object segmentation purpose [21][22][23][24]. However, these semantic segmentation techniques have not extensively been explored for malaria parasites segmentation in thin blood smears. On the other side, the limitation of semantic segmentation techniques is primarily raised since the large object classes dominate the segmentation task, and then the small object classes are usually suppressed. This condition often provides unsatisfied performance of object detection on small objects [25].

Another challenge is collecting a dataset. Dataset is a crucial part of developing CAD systems. Data training has a significant influence on a model during predicting data testing. Therefore, the collection of datasets from the condition of the system to be implemented is essential. Some public microscopic malaria datasets have been offered, such as follows. Dong et al. [26] extracted cell patches from a larger malaria dataset created by the University of Alabama at Birmingham (UAB). The cell patches contain two classes, namely infected and non-infected. Both classes consist of 1,034 and 1,531 cells, respectively. However, the type of parasite on this dataset did not describe.

Rajaraman et al. [27] captured 200 thin blood smear microscopic images using a smartphone camera at Chittagong Medical College Hospital, Bangladesh. The dataset consists of 50 non-existing and 150 existing parasites. An expert slice reader at the Mahidol-Oxford Tropical Medicine Research Unit manually annotates the dataset for parasite detection and patch classification tasks. However, the infected images only contain parasites of falciparum.

Yang et al. [28] collected a malaria dataset captured by a camera smartphone on thick blood smears. The dataset contains 1,819 captured from 150 subjects. They proposed malaria parasite detection in two-stage. The first is generating malaria parasite candidates using Iterative Global Minimum Screening (IGMS) based on intensity pattern. The second is eliminating the artefacts from malaria-generating candidates by customized CNN classification.

Segmented-malaria dataset compiled by [29] offered a public dataset for segmentation tasks. The dataset contains 27,558 images of individual cells; which experts manually annotate the contour of the parasites. However, this dataset contains one parasite type, falciparum, and only individual images of blood cells which are unusual to implement in the real world.

Indonesia is the largest archipelagic country having uneven economic development. Unfortunately, common malaria cases occur in Indonesia's eastern areas with low economic development. Consequently, the smear collection was carried out with minimal health facilities, vulnerable to smear damage during smearing and storage. This condition causes some smears to contain many artefacts, such as dust. However, regarding the offered public dataset, those are minimized from artefacts. Therefore, in this study, we collect a dataset according to natural conditions and without choosing only good images as a fundamental step to developing a CAD system for our country.

Regarding that problems, this study provides two significant contributions:

1.  This study collects an extensive malaria parasite dataset originating from Indonesia called the PlasmoID database. Most of the data are obtained from rural malaria areas. The database comprised all parasite types with their stages and their ground truth for detection and segmentation purposes.
2.  As a preliminary study, this study also proposes a new scheme for malaria parasite detection and segmentation by combining the faster RCNN and semantic segmentation techniques to advance the performance of parasite segmentation and detection. The proposed scheme is evaluated on PlasmoID database and compared with their original semantic segmentation techniques.

## 2. Related Works

Several recent studies in parasite detection and segmentation techniques based on a deep learning approach have become popular because they have good performance and reliability. The techniques are explored as follows.

Fuhad et al. [30] proposed a new model for detecting malaria parasites on segmented RBC patches. They evaluated the proposed model on their public dataset containing 27.558 cell images. As a result, their model achieved more than 99% accuracy. However, this study works on a segmented cell image that is unusual applied in the real world. The others similar study tasks are [31][32][33].

Sifat et al. [34] suggested an automated system to detect infected RBCs and classify them into parasite classes using deep learning approaches. Three main proposed steps are infected RBC candidate segmentation using UNet, infected and Uninfected RBC classification using CNN, and malaria parasite classification using VGG16. They evaluated their model on a large public dataset [35]. However, they only informed the model segmentation performance only accuracy and specificity, which both of them involve true negative (background). Therefore, the performance is high.

Delgado-Ortet et al. [36] introduced malaria detection through three steps. First is segmenting (red blood cells) RBCs. The second is cropping and masking the RBCs, and finally, RBCs patches classification. The first and third steps apply a deep learning approach for segmentation. This study evaluated their model in two datasets; one of them is a public dataset. However, the type of parasites contained in their dataset is not informed. Nevertheless, the model performance accuracy of segmentation achieved 93.7%.

In a recent study, Loh et al. [37] introduced malaria cell infection detection and segmentation using a deep learning approach, Mask-RCNN. Generally, their method through two steps, region proposal and region segmentation. The model is trained on uninfected and infected RBCs. Referring to their results, their model achieves high accuracy of 94.6% and is fast. However, this article did not inform the number of microscopic images, and the data only contains malaria falciparum.

Previous studies above investigated malaria detection using RCBs detection or segmentation in their early stage. Their dataset offered a ground truth for RBC detection and segmentation. It indicates that the RBCs individually occur on their microscopic images. However, in another dataset case, RBCs overlapped. Therefore, the common strategies cannot be applied to these datasets and our dataset. Some previous works that presented a solution to this case are listed below.

Recently, Adrien et al. [38] performed UNet combined with the Green Green Blue (GGB) image normalization technique to segment malaria parasites in thin blood smear microscopic images. They compared several colour spaces in the input image, including

RGB, HSV and GGB. The proposed achieved the best segmentation accuracy of 99.47%. However, they only calculated accuracy segmentation performance, and the data used was public contained 222 images which were minimum noise and artefact.

Abraham [39] also applied UNet to segment malaria parasite patches with the Green Green Blue (GGB) image normalization technique to segment malaria parasites in thin blood smear microscopic images. In exploring the best combination of loss function performance, this study combined three loss functions: mean squared error, binary cross-entropy, and Huber loss. Huber loss achieved the best performance in segmenting parasites with 93%, 97.5%, 89.6% and91% for F1-score, PPV, sensitivity, and relative segmentation accuracy (RSA). Unfortunately, the data contained only 30 patches sized 200x200 pixels in one parasite species, Falciparum-Gametocyte. The other study applying UNet in microscopic images was conducted by Gorriz et al.[40]. They applied UNet to segment leishmania parasites and classify them into three classes, namely, amastigotes, promastigotes, and adhered parasites. The performances in some classes achieve a satisfying result.

Most previous studies that worked on parasite segmentation applied UNet architecture. Other newest architectures for semantic segmentation have not been explored for parasite segmentation. He et al. [22] proposed ResNet architecture that aims to solve the deep gradient degradation problem. His model won first place in the ImageNet competition classification task. The Res-UNet runs by combining U-Net and ResNet with some modifications. The modification processes are described as follow. (1) The convolutional layer, pooling layer, and residual unit were designed by adopting the basics of ResNet. (2) Residual concept inspired by ResNet was applying as a feature extractor in the downsampling layer and the upsampling layer. (3) Linear interpolation techniques applied in deconvolution step. (4) The last, the number of final output classes is adjusted according to user requirements [41].

One of improved FCN's was published in [23]. The improved FCN is called ResFCN-18. The model has been applied to segment the Steel bar end face. The input image of FCN is a large size. Three times of down-sampling of the feature map and deconvolution layer are required to produce small image patches. The feature map is an essential part of extracting deep object features. They put three feature extractor models; one of them is ResNet 18.

The last recent segmentation technique based on the deep learning approach is DeepLab V3 [24]. This technique was known for the different concepts of feature map extractors, namely Atrous convolution. The convolution concept has an advantage compared to the traditional. Traditional convolution and pooling make the output stride increase, i.e. the output feature map smaller when extracting deeper. Therefore, this approach is not recommended for semantic segmentation because some spatial information is lost at the deeper layers.

In this study, we explore some deep learning architectures in instant segmentation applications to segment malaria parasites in our challenging dataset. However, semantic segmentation techniques have a limitation in segmenting the small objects when the large object classes dominate. Therefore, some improvements are required in order to segment the small object accurately. In this paper, we also propose a new strategy to advance the detection and segmentation of malaria parasites by combining semantic segmentation and object detection techniques. The proposed scheme is evaluated in the PlasmoID dataset. In addition, we applied the proposed scheme to five recent semantic segmentation models and compared them to their original ones. These comparisons aim to show our strategy's reliable performance to enhance the accuracy of the five compared models.

## 3. Materials and Methods

### 3.1. Collection of dataset

This study collects an extensive dataset of malaria parasites originating from Indonesia. The dataset contains 559 digital microscopic images consisting of 91 negative malaria images and 468 infected malaria images captured on thin blood smear at 1000 times magnification level. The infected malaria images contain 691 total parasites. We use an Optilab digital microscope camera with an output size of 960x1280 pixels to capture the thin blood smear. The smear is collected from the Eijkman Institute and the medical faculty of UGM, the majority of which came from rural Indonesia.

The collected dataset contains 12 classes. The classes comprise four types of malaria parasites with their three life stages. The malaria parasites are P.Falciparum, P.Vivax, P. Malariae, and P.Oval, and their life stages are Trophozoite, Schizont, and Gametocyte. An experienced microscopic examiner and a parasitology were invited to assist capturing the images and confirm the parasites' type, life stage, location, and boundary. In this study, the collected parasite database is randomly divided into 0.8:0.2 for training and testing purposes. Fig. 1 shows examples of malaria parasite morphology in every class.

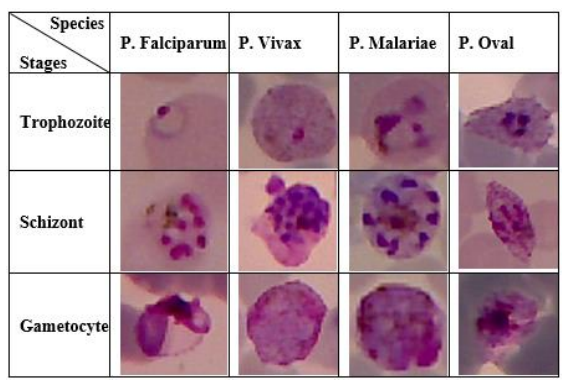

**Figure 1.** Several malaria parasite species with their life stages of PlasmodID dataset.

### 3.2. Proposed Scheme

Image segmentation techniques are a significant step in automated computer-aided diagnosis, especially in medical, since they provide more detailed object features for more profound decisions than location detection techniques, even image classification techniques. Some semantic segmentation techniques based on deep learning have recently become popular due to their excellent performance and fast. However, the common problem of several semantic segmentation techniques is raised when large objects are dominant. Hence, it generally causes unsatisfactory results in detecting small objects.

In addressing the limitation of the semantic segmentation technique, this study proposes a new strategy using a combination of object detection and a semantic segmentation technique. Our scheme has three main steps. The first step is applying a semantic segmentation technique. In the second step, we apply an object detection technique to detect the parasite candidate locations and a semantic segmentation technique to segment the selected parasite candidate patches. We use a rule-based strategy to determine the patches. Finally, we combine the first and second steps using a rule-based approach. The detailed procedure will be discussed as follow. In general, the diagram block of our proposed scheme is depicted in Fig. 2.

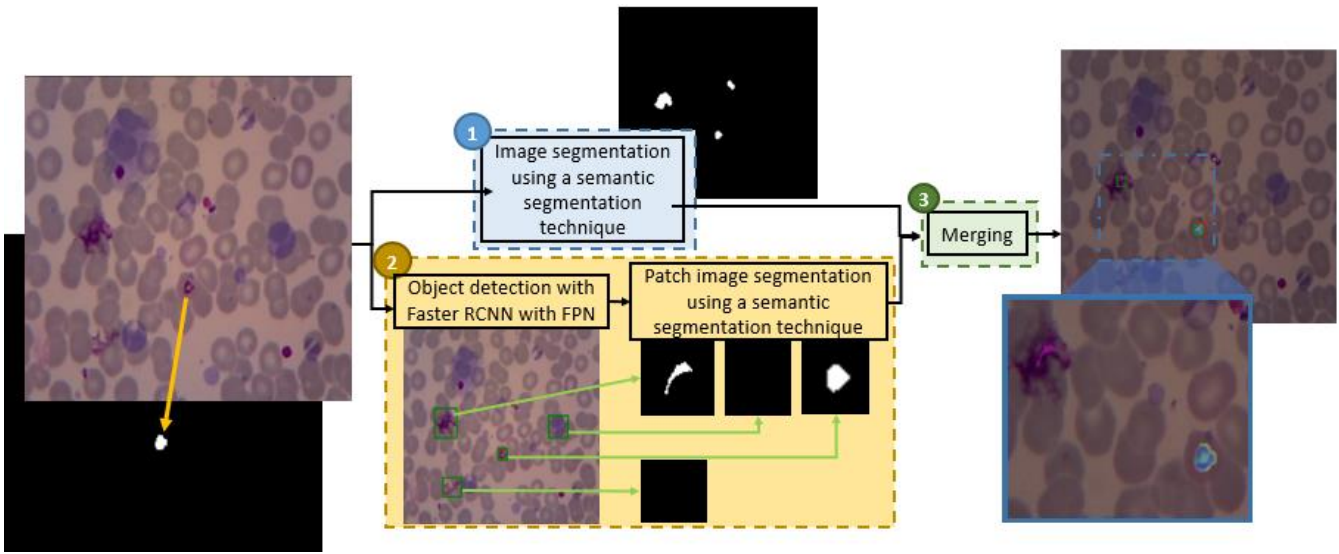

**Figure 2.** Block diagram of our proposed scheme.

First, image segmentation step aims to segment the whole image. In this step, we use a recent semantic segmentation algorithm. Second, object detection step aims to detect all object potentially being parasites. In this step, we apply a combined Faster-RCNN with FPN [42]. The diagram block of the combined Faster-RCNN with feature pyramid network (FPN) is shown in Fig. 3. FPN is a feature extractor that generates feature maps in several image scales (low to high-level scales). Therefore, the design of FPN is like a pyramid feature map. Applying FPN aims for the model to learn more about the input image's details. Each extracted feature map feeds to a region proposal network (RPN) to continue the process, as in the Faster-RCNN algorithm. Referring to the parasite candidate locations with a size lower than RBC ($\alpha$), we create a square patch as RBC's size. This approach aims to get the background of RBC. Therefore, in the following object detection step, patch segmentation, the model can accurately predict background, artefact or parasite intensity. $\alpha$ is obtained by averaging the size of parasites on the training images. In segmenting the predicted object, we use a semantic segmentation technique in which the architecture is the same as in the first step. Third, merging the previous two stages aims to reduce the false positive (FP) rate. We use AND logic to eliminate FPs object detection in the first and second steps. The merging step is applied only to the predicted objects that have over $\alpha$. For the object sized under $\alpha$, we use the second step to segment the predicted objects. Therefore, the output of our proposed scheme is the segmented image. Our scheme will be applied to five recent semantic segmentation models and compared with their original ones to show the reliable performance of our scheme.

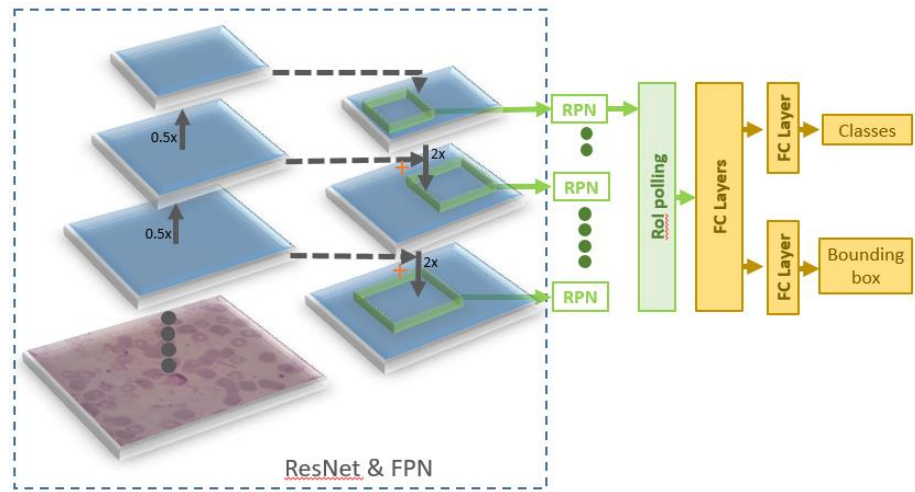

**Figure 3.** Diagram block of combined Faster-RCNN and FPN using ResNet structure [42].

### 3.3. Simulation procedures and controls

Referring to the sub-path of the proposed method, we train the three architectures in three steps. The steps are semantic segmentation architecture for image segmentation, detection architecture for small object detection, and semantic segmentation architecture to segment the small predicted patches obtained by the object detection step. Hence, we use three different training procedures explained as follow.

Each model's input in all steps required a tensor format with the size of Batches x Height x Width x Colourspace. The size needs to be adjusted to meet the image quality with the capabilities of our device. We also apply the Gaussian filter technique to avoid aliasing artefacts when down-sampling or resizing an image. Each model is trained on 80% of the total data and tested on the rest. We use a transfer learning strategy to accelerate the training process to reach a steady state.

#### 3.3.1. Image segmentation step

We apply five semantic segmentation algorithms for performance comparison namely, FCN-res18 [23], DeepLabv3 mobilenetv2 [24], DeepLabv3plus mobilenetv2 [24], UNet-base [21], and ResUNet-18 [41]. In the training phase, the input model is adjusted to size 2x512x512x3, and the number of classes is set to one. At this step, we use image normalization to follow the training model's previous procedure on ImageNet. The detail of the image normalization technique is explained in [43]. We use the Adam optimization technique during the training phase with the initial learning rate (LR) of 0.003 and the learning rate decay of each parameter group with a gamma of 0.1.

A light augmentation technique containing six image processing algorithms, namely, horizontal flip, vertical flip, Affine-scale, Affine-translate, Affine-shear, and Affine-rotate, is used in this step. One of horizontal flip and vertical flip is randomly chosen. The others are applied to half of all images that are randomly chosen. In addition, all parameters of these algorithms are also randomly adjusted.

#### 3.3.2. Object detection step

We set the input model as a tensor sized 2x512x512x3. We use faster-RCNN combined with FPN to detect the parasites. The number of object detection classes is one, i.e. parasites. In addition, we use the SGD optimization technique to optimize the training with the initial learning rate of 0.0005. We also use a learning rate scheduler that decays the learning rate by a factor of 0.1 every three epochs.

A custom light augmentation technique is applied. The augmentation design is similar to the image segmentation step. Horizontal flip and vertical flip are applied on 50% and 20% of all images.

### 3.3.3. Candidate patches segmentation

This step aims to segment the candidate patches obtained by the object detection step. The size of the input model is adjusted to 2x128x128x3. The adjusted size is the closest eight to the power of n to the average parasite size. At this step, we also use image normalization and learning optimization as in the first step.

We also apply a light augmentation design to enrich the image patches. This design augmentation concerns enhancing the variance of texture. The augmentation contains five image processing techniques, namely, horizontal flip, vertical flip, sigmoid contrast, log contrast, and linear contrast. One of horizontal flip and vertical flip and one of the others are randomly chosen.

### 3.4. Evaluation

This study offers a dataset for malaria parasite detection and segmentation from Indonesia named PlasmoID dataset. The other offer is that we proposed a scheme for malaria parasite detection and segmentation on the PlasmodID dataset. The proposed scheme desired to enhance some semantic segmentation models' parasite detection and segmentation performances. We applied the proposed scheme to five recent semantic segmentation models and compared them to their original ones. These comparisons aim to show our strategy's reliable performance to enhance the accuracy of the five compared models. The detailed evaluation procedures are described below.

The compared techniques are evaluated in the PlasmoID database. However, before we are going forward, some basic evaluation definitions must be explained as followed. The output is either a parasite pixel (positive) or a background pixel (negative), only two classes of segmentation class.

- True positive (*TP*): Prediction is positive, and X is a parasite pixel; we *want that*
- True negative (*TN*): Prediction is negative, and X is a background pixel; we *want that too*
- False-positive (*FP*): Prediction is positive, and X is a background pixel; *false alarm, bad*
- False-negative (*FN*): Prediction is negative, and X a parasite pixel; *the worst*

These rules are illustrated in Fig. 4. This study aims to segment parasites on the microscopic image, where the parasite area is much smaller than the background. Therefore, we exclude the evaluations that contain TN to prevent the evaluation parameters from having a high value. The remained evaluations are sensitivity or recall and precision or PPV. The correlation between sensitivity and precision is F1-score which is formulated in Eq. 1.

| | | **Predicted Class** | | |
|---|---|---|---|---|
| | | **Positive** | **Negative** | |
| **Actual Class** | **Positive** | True Positive (TP) | False Negative (FN) *Type 1 Error* | **Sensitivity** $\frac{TP}{(TP + FN)}$ |
| | **Negative** | False Positive (FP) *Type 1 Error* | True Negative (TN) | **Specificity** $\frac{TN}{(TN + FP)}$ |
| | | **PPV** $\frac{TP}{(TP + FP)}$ | **NPV** $\frac{TN}{(TN + FN)}$ | **Accuracy** $\frac{TP + TN}{(TP + TN + FP + FN)}$ |

**Figure 4.** Matrix confusion and performance calculations.

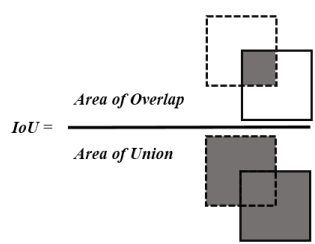

**Figure 5.** Illustration of IoU calculation.

*F1 score = 2\* (Precision \* Recall) / (Precision + Recall)*          (1)

For object detection evaluation, we should determine how well the system predicts the location of all the parasites. The localization is evaluated by calculating the overlap ratio between the ground truth bounding box and the predicted bounding box. The overlap ratio was called intersection over union (IoU). The IoU illustrates in Fig. 5. To evaluate the predicted location, we need to set the IoU threshold value. In this study, we use the IoU threshold value of 0.3. This threshold was recommended by the previous studies [20][19].

## 4. Result and Discussion

### 4.1. Dataset collection

The total collected data is 468 infected malaria images, with 691 total parasites and 91 negative malaria images, most of which came from rural Indonesia. We also provide three ground truths for each parasite, namely localization, segmentation, and classification that had been confirmed by an experienced microscopic examiner and a parasitologist. The data is called the PlasmoID database.

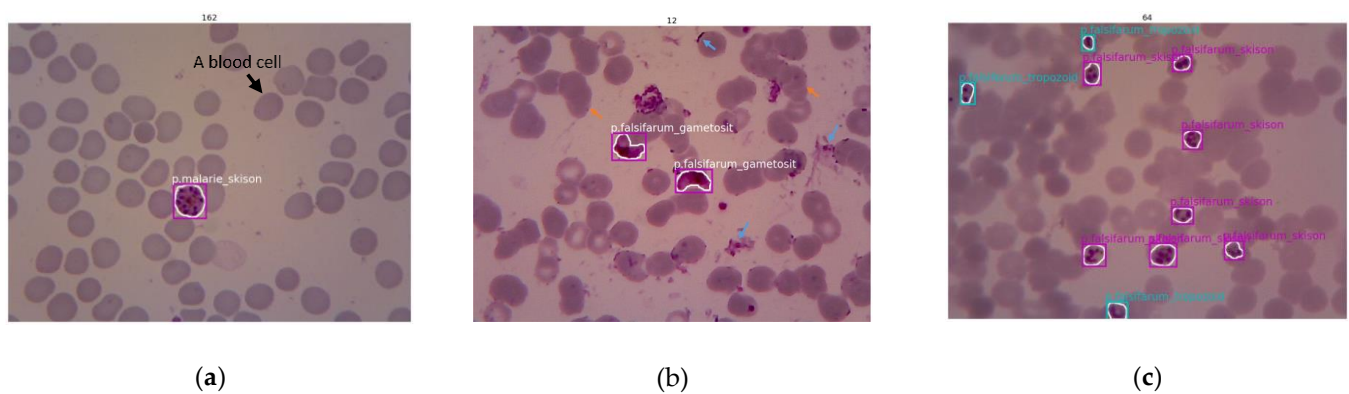

| (a) | (b) | (c) |

**Figure 6.** Examples of the microscopic images in the PlasmoID database with their ground truth, (a) a good microscopic image with minimum artefact and noise, (b) an image containing some parasites and artefacts, (c) image containing many parasites of different classes, and blurred red blood cells.

Fig. 6 shows some thin microscopic images on the PlasmoID dataset. Magenta boxes indicate the ground truth bounding box of parasites. In addition, the white contour represents the ground truth segmentation. Fig. 6(a) is one good example of a photo on the PlasmoID dataset. However, the PlasmoID dataset mostly contains photos having many artefacts such as dust, platelet, blur, and luminance noise. Fig. 6(b) is a photo containing multiple parasites with some artefacts. The blue points indicate the artefacts, and the orange points indicate the joined red blood cells. The parasites commonly appear within a single class in a microscopic image. Fortunately, we have the microscopic images that contained multiple parasites within multi-classes, as shown in Fig. 6(c). In addition, blurred red blood cells also appear in this photo.

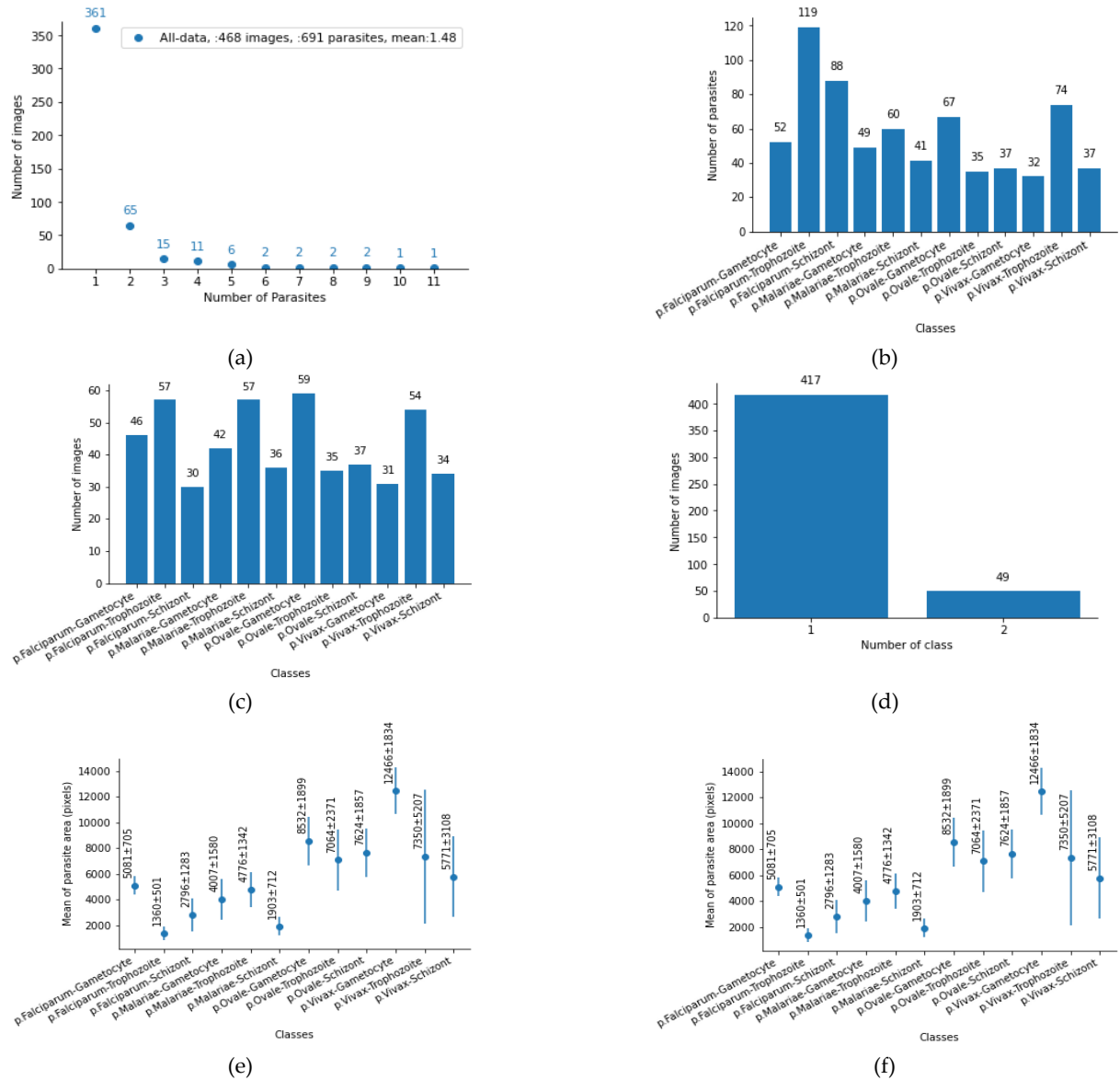

**Figure 7**. The PlasmoID data distributions, a) the distribution of parasites in all parasite classes, b) the distribution of parasites in each parasite class,  c) the distribution of images in each parasite class, d) the distribution of images in the number of parasite classes, e) the distribution of mean area parasites in each parasite class, f) the distributions of parasite mean's width and height in each parasite class.

The parasite distribution on the infected images is depicted in Fig. 7(a). The graph shows that more than 350 of 468 images only have one parasite with a mean distribution of 1.48 parasites/image. Fig. 7(b) show the distribution of parasite in each class. The highest number of parasites is in *P. falciparum gametocyte* (119 parasites), and the lowest is in *P. vivax gametocyte* (32 parasites). The unbalanced data is caused by the uneven distribution of malaria cases in each class in Indonesia. Fig. 7(c) shows the distribution of the number of images in each parasite class. *P. falciparum trophozoite, P. malariae schizont, P. ovale gametocyte*, and *P. vivax trophozoite* are the commonly-occurring parasite classes. The PlasmoID dataset has some photos containing multi parasites and also multi parasite classes, such as shown in Fig. 6(c). The distribution of the number of photos containing multiple parasite classes is shown in Fig. 7(d). Fig. 7(e) and 7(f) show the distribution of the average parasite area and parasite bounding box for each parasite's class.

The collected data and the ground truth are stored in .png and .json formats. PlasmoID provides image specifications and ground truth in each image, namely width, high, ID, and annotations. The width and high are the original image size. The ID contains an address and the image file name. Finally, the annotation contains parasite ground truth (GT), whose GT number depends on the number of parasites in the image. Thus, every GT contains the GT of parasite detection ('bounding_box'), parasite segmentation (coor_H and coor_W) and its class (parasite type and stage). This study uses a data division of 0.8 for training and 0.2 for testing. This study aslo use only use sigle class namely parasite class. The distribution of the divided data is shown in Fig. 8.

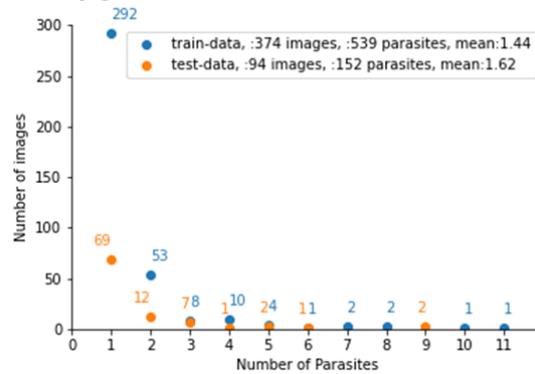

**Figure 8**. The distribution of parasite in training data and testing data.

### 4.2. Parasite segmentation and detection

This study develops a scheme for malaria parasites segmentation and detection in thin blood smears. The proposed scheme desired to enhance some semantic segmentation models' performance. We compare the performance of a semantic segmentation architecture combined with our scheme (proposed) and without our scheme (original) on five recent semantic segmentation architectures to show the reliable performance of our scheme. The architectures are FCN-res18 [23], DeepLabv3 with backbone mobilenetV2 [24], DeepLabv3plus with backbone mobilenetV2 [24], UNet-base [21], and ResUNet-18 [41]. Table 1 shows the performance comparison of segmentation methods. The blue bars indicate the performances of five original semantic segmentation techniques. The orange bars represent our proposed schemes combining with their semantic segmentation techniques.

**Table 1.** Segmentation performance comparison of the proposed scheme and the original semantic segmentation methods.

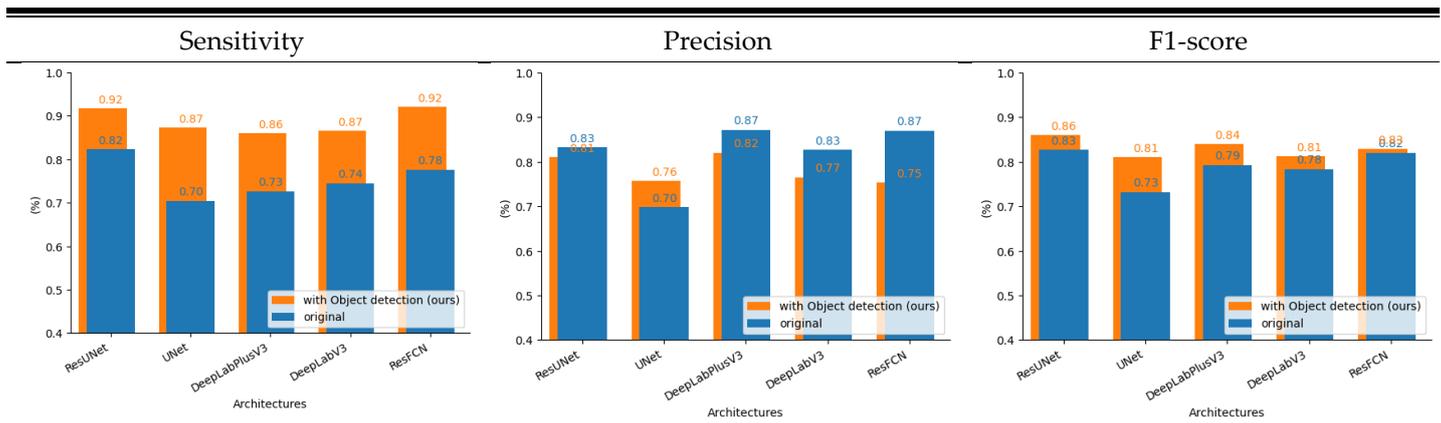

Referring to the Table 1, the proposed scheme can improve sensitive and F1-score performances of all semantic segmentation techniques for segmenting malaria parasites in the PlasmoID dataset. Decreasing the precision performance in four techniques is due to inaccurate segmenting of the small predicted patches. Inaccurate is caused by insufficient information while training their model since the input patch size was tiny. However, on the other hand, our scheme increases sensitivity and F1-score performances. This advantage is caused by applying an object detection technique to find small objects. Finding small objects was a limitation of the semantic segmentation technique when the large objects were dominant. The success of applying the object detection technique is proven in Table 2. By using our scheme, all compared semantic segmentation techniques have increased the performance in all object detection evaluation parameters.

**Table 2**. Detection performance comparison of the proposed scheme and the original semantic segmentation methods.

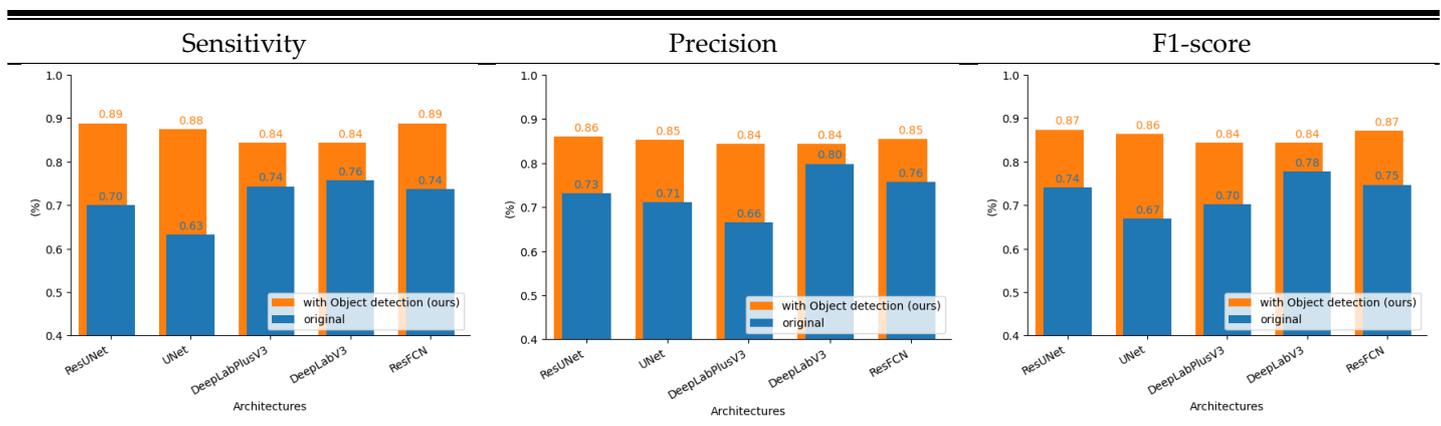

Fig. 9 and 10, the comparison results between the proposed scheme and the original techniques, have three different mask colors. The mask with magenta, yellow and cyan are under-segmentation, over-segmentation and the correct segmentation. Fig. 9(a) shows the segmentation results of ResUNet-18, which has four small uncovered/ undetected parasites. Unsatisfactory segmenting of the small objects is when the large objects are dominant, which is the main limitation of semantic segmentation techniques. Fig. 9(b) is the result of our proposal, ResUNet-18 combined using our scheme. By applying our proposed scheme, the undetected objects are remained only one.

Regarding to the Table 1 on the precision column, the proposed scheme, combining ResFCN-18 with Faster-RCNN, has the highest drop performance. The performance degradation is caused by insufficient of the input patch size while in the training process. Fig 10 (b) show that our proposed scheme successfully detects the small parasites better than the original ResFCN-18, but over-segmentation most occurs on small objects. The other limitation of our proposal is the computational time. After combining three models in object detection and semantic segmentation, the computational time increased. However, the average increase in computational time is only about 0.2 from 0.19 sec/image. Therefore, it has no significant effect when our proposed scheme is applied to CAD to detect and segment malaria parasites.

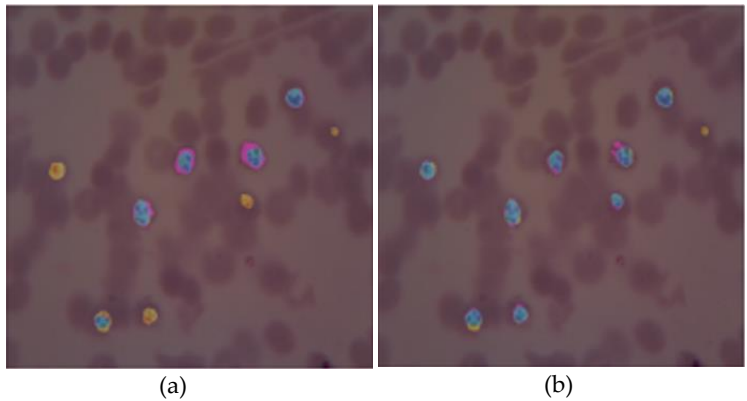

**Figure 9**. A comparison result of parasite detection and segmentation. a) ResUNet-18, and b) our proposed scheme (combination ResUNet-18 and Faster-RCNN).

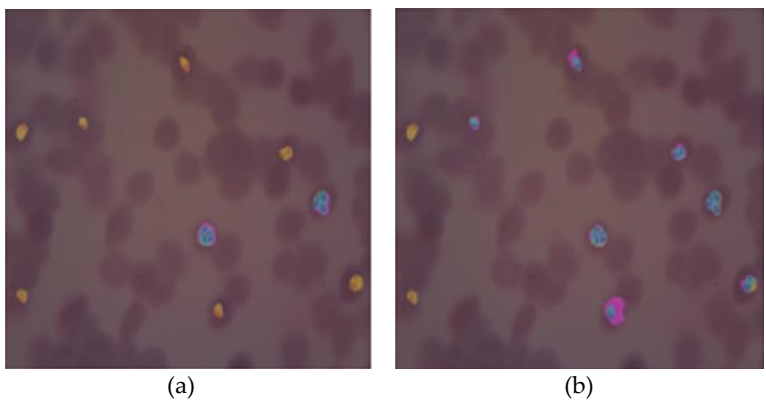

**Figure 10**. A comparison result of parasite detection and segmentation. a) ResFCN-18, and b) our proposed scheme (combination ResFCN-18 and Faster-RCNN).

## 5. Conclusion

We conduct two main contributions in this study. First, we collect the malaria parasite dataset originating from Indonesia. The dataset contains 559 microscopic images of thin blood smears with 691 total parasites on all their species. The dataset is called PlasmoID. The PlasmoID also provides the ground truth (GT) for parasite detection and segmentation purposes. The GT has been validated by an experienced microscopic examiner and a parasitologist. Second, we propose a new scheme as preliminary study on PlasmoID dataset for malaria parasite detection and segmentation by combining semantic segmentation technique and object detection technique. The proposed scheme desired to enhance some semantic segmentation models' performance. We compare the performance of a semantic segmentation architecture combined with our scheme (proposed) and without our scheme (original) on five recent semantic segmentation architectures to show the reliable performance of our scheme. Our proposed scheme has better performances in sensitivity and F1-score for malaria parasite segmentation and has better all performance for malaria parasite detection compared to the five compared techniques. These indicate that this study is essential in developing automated parasite detection and segmentation in thin blood smears, especially in the PlasmoID dataset.

## Acknowledgement

The authors would like to acknowledge the Department of Parasitology, Faculty of Medicine, Public Health and Nursing, Universitas Gadjah Mada and Eijkman Institute for providing the database in this research and also to The Deputy of Research and De-velopment, National Research and Innovation Agency Republic of Indonesia through the Research Grant of "World Class Research" for supporting the funding in this research.